\documentclass[a4paper,11pt]{article}
\usepackage{pos}
\usepackage{ulem}

\usepackage{lineno}

\title{The NA60+ experiment at the CERN SPS to study dilepton and heavy quark production at large $\mu_{B}$}
\ShortTitle{The NA60+ experiment}

\author*[a, b, 1]{Maryna Borysova}
\note{For the NA60+ collaboration}

\affiliation[a]{Weizmann Institute of Science,\\
  234 Herzl St., Rehovot, Israel}

\affiliation[b]{Institute for Nuclear Research,\\
  47, prospekt Nauky, Kiev, Ukraine}

\emailAdd{maryna.borysova@weizmann.ac.il}

\abstract{The region of high baryonic densities of the QCD phase diagram is the object of several studies focused on the investigation of the order of the phase transition and the search for the critical point. The rare probes, which include electromagnetic observables and heavy quark production, are experimentally challenging to access as they require large integrated luminosities that could be studied with fixed-target experiments. A future experiment, NA60+ at CERN, is being proposed to access this region and perform accurate measurements of the dimuon spectrum up to the charmonium region and study charm and strange hadrons. With its high beam intensity, the CERN SPS can cover the center-of-mass collision energy region from 6 to 17 GeV providing access to rare observables which have been scarcely studied until now. The proposed experiment includes a muon spectrometer based on tracking gas detectors coupled to a vertex spectrometer based on Si detectors. The time slot after the Long Shutdown 3 of the LHC (past 2029) is aimed for the first data-taking, with Pb and proton beams.
In this contribution, we review the project and recent R\&D effort, including the technical aspects and the studies of the physics performances for the observables.
}

\FullConference{%
  41st International Conference on High Energy Physics - ICHEP2022\\
  6-13 July, 2022\\
  Bologna, Italy
}

\begin{document}
\maketitle

\begin{figure}[ht]
\begin{center}
\includegraphics[width=0.5\linewidth]{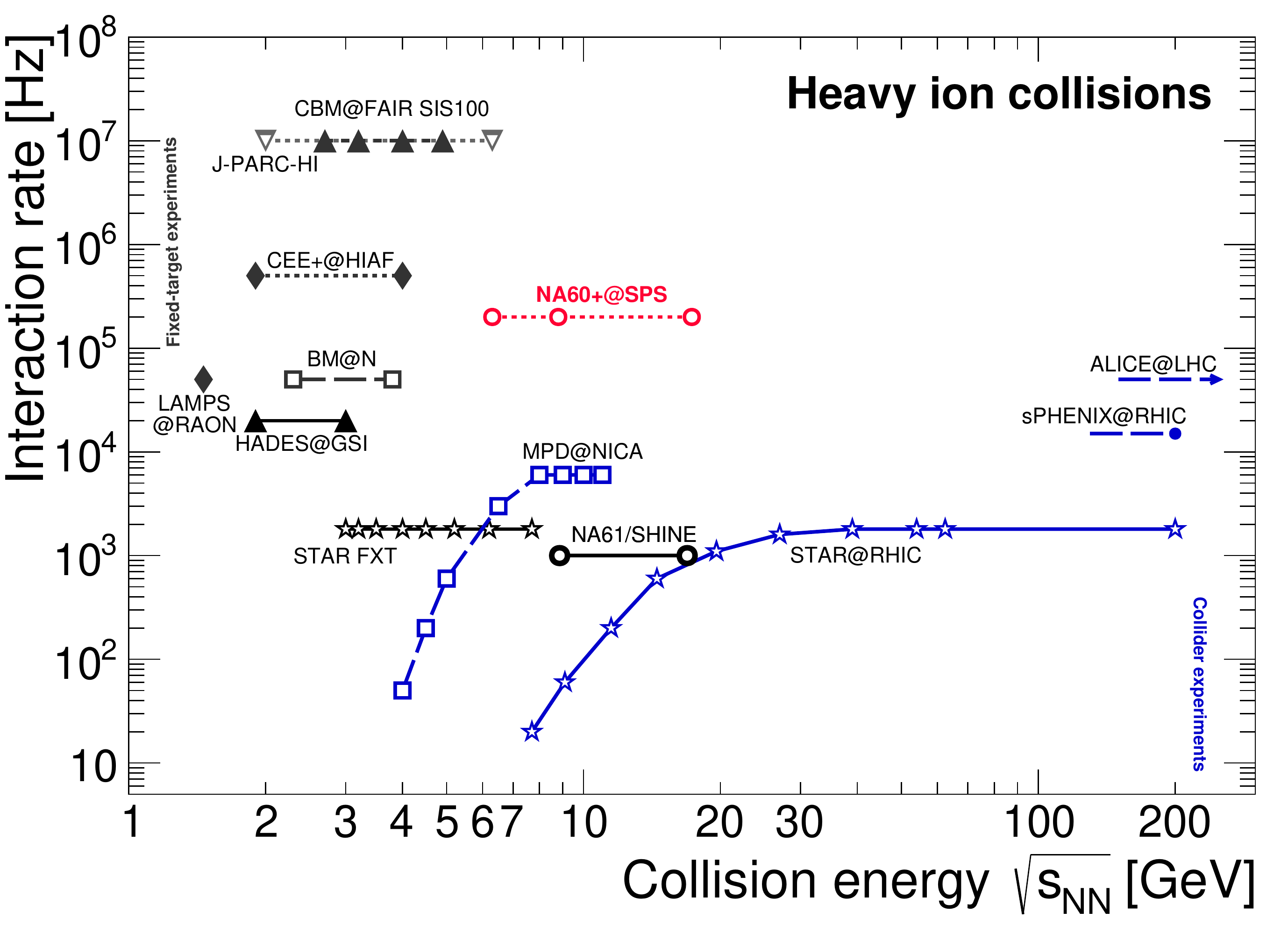}
\caption{Existing and planned nuclear beams experiments~\cite{Galatyuk:2019lcf}.
%The collider experiments are highlighted in blue, and fixed-target experiments are in black. The NA60+ parameters are shown in red.
}  
\label{fig:Tetyana}
\end{center}
\end{figure}

%\section{Introduction}
NA60+, a successor of NA60, is a new dimuon experiment that aims to study hard and electromagnetic processes at CERN-SPS energies. It extends and improves the physics program of its predecessor by performing measurements of dileptons and heavy-quark production over the entire SPS energy range and increasing the precision of the NA60 results by using an improved apparatus based on cutting-edge technologies. The NA60+ project will be part of a worldwide experimental program studying the properties of the Quark-Gluon Plasma (QGP) in the region of relatively large baryochemical potentials ($\mu_{\rm B}$). 
%These studies require much lower center-of-mass energy than the top energy that can be reached at the RHIC and LHC ion colliders. 
Figure~\ref{fig:Tetyana} 
shows current and foreseen experiments in terms of energy coverage and interaction rate for nuclear collisions~\cite{Galatyuk:2019lcf}. 
In the SPS colliding energy range, which covers the interval between 6 and 17 GeV per nucleon in the center-of-mass frame, the NA60+ experiment, with a foreseen interaction rate exceeding $10^5$\,s$^{-1}$ has a unique position %in parameters space in terms of
reaching counting rates up to two orders of magnitude higher than other experiments. 
%This is important for studying rare processes. 
At these energies, NA60+ can explore the range  $230<\mu_{\rm B}<560$\,MeV where the QCD phase diagram should have a first-order phase transition between hadronic matter and QGP which terminates at a second-order critical point. Discovering signals of the first-order phase transition and the location of the critical point represents one of the hottest topics of relativistic heavy-ion physics. 

\begin{figure}[ht]
\begin{center}
\includegraphics[width=0.8\linewidth]{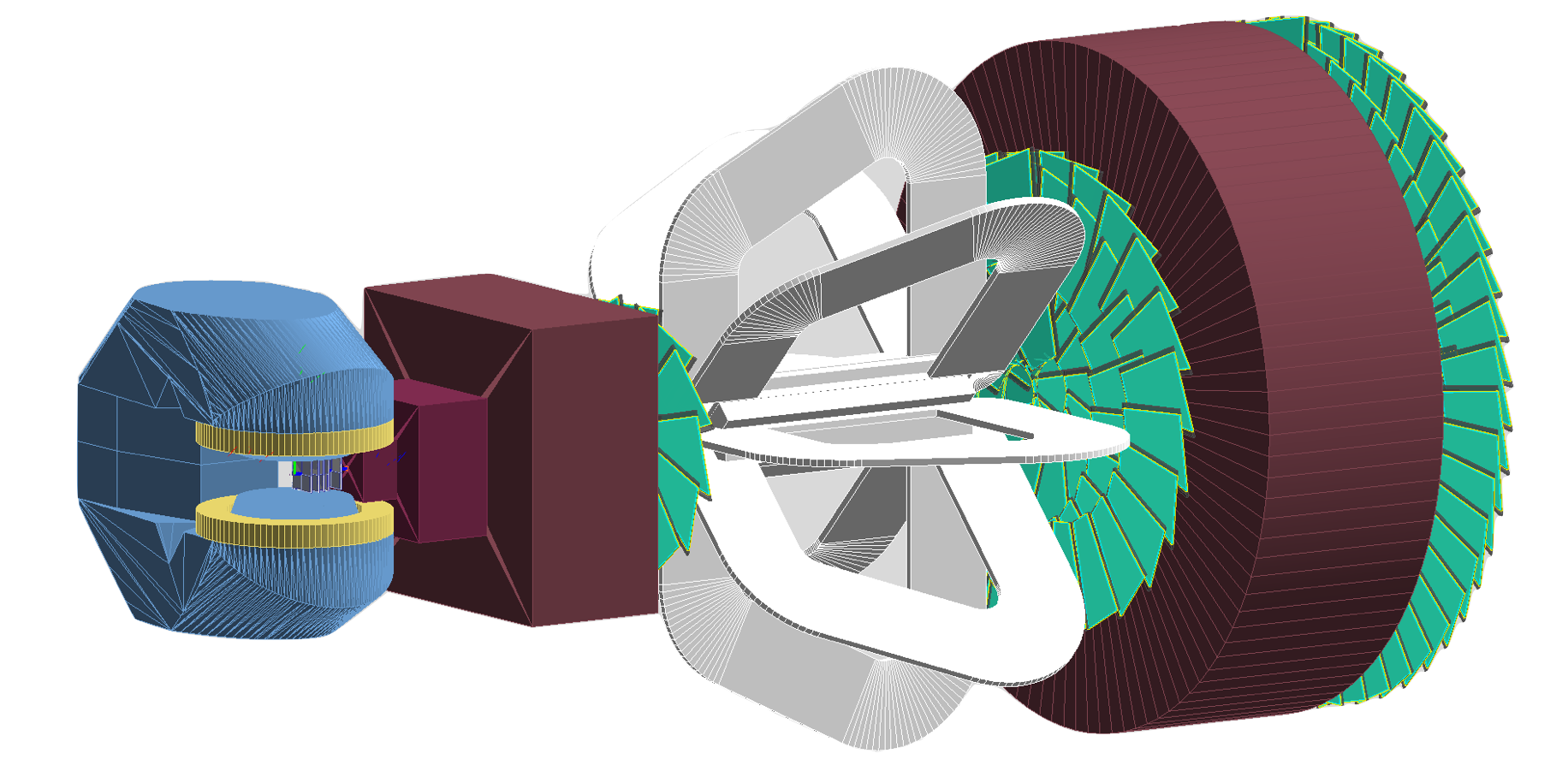}
\caption{GEANT4 rendering of the NA60+ experimental apparatus optimized for low-energy data taking. A C-shaped dipole magnet is shown in blue/yellow. Inside the dipole in light blue are the vertex spectrometer planes. Muon tracking chambers are shown in green, the purple blocks are absorbers, and the toroid magnet is grey.
}
\label{fig:NA60concept}
\end{center}
\end{figure}

%\section{The NA60+ concept}
Significant improvement and extension of the physics reach with the NA60+ apparatus will be possible due to a high-intensity beam with at least 10$^7$ Pb ions per spill and state-of-the-art experimental technologies. The setup of the NA60+ experiment is shown in Fig.~\ref{fig:NA60concept}.
It consists of several elements. A target system composed of several sub-targets, possibly including various nuclear species, is foreseen.
%Five relatively thick targets can be made of various elements. Their thicknesses are chosen to increase the collision rates. Together with the beam, they define the system of colliding species. 
Targets are immediately followed by the vertex spectrometer immersed in a 1.5\,T magnetic field of a dipole magnet MEP48, providing a field integral of about 0.5\,Tm. A vertex spectrometer, consisting of 5 (up to 10) identical silicon pixel planes, is positioned at $7<z<38$\,cm starting from the most downstream target. Each plane, featuring a material budget of 0.1\% X$_0$ and intrinsic spatial resolution of $\sim$5 $\mu$m,  is formed by four large area monolithic pixel sensors (MAPS) of 15x15\,cm$^2$ each.
%An expected momentum resolution of the vertex spectrometer allows reaching the invariant mass resolution of ... MeV at ... GeV.
An absorber made of BeO with a fixed length of 105\,cm is followed by a graphite absorber that is at least 130\,cm thick, in the version of the set-up optimized for low-energy data taking. The absorber is positioned immediately downstream of the vertex spectrometer, in order to reduce the probability of K and $\pi$ decays into muons and suppress combinatorial background. The choice of the material is dictated by requirements of relatively high density and low Z to minimize multiple scattering of muons. The center of the absorber is made of tungsten, in order to dump the non-interacting ions and forward-emitted fragments. One of the features foreseen in the NA60+ setup is a variable length of the absorber in order to comply with the increase of hadron multiplicity with collision energy.
This can be done by increasing the graphite section up to 335\,cm.
The increase of the absorber thickness also implies that the consequential parts have to be moved downstream, ensuring the rapidity coverage to remain constant with varying collision energy, in the approximate range $0<y<1$.
%This is done to ensure the rapidity coverage to be more than one unit near center-of-mass rapidity at all beam energies while simultaneously providing more absorption length for the more energetic pions and kaons. 

Muons traversing the absorber lose about 1\,GeV of their energy. They are analyzed in the Muon Spectrometer (MS), which is built out of two stations of pad detectors separated by a toroidal magnet with a 0.5\,T magnetic field. Pad chambers will be based either on Multi-Wire Proportional Chambers (MWPC) or Gas-Electron Multipliers (GEMs) technique. Their goal is to provide approximately 200\,$\mu$m spacial resolution while handling a maximum flux of particles of the order of a few cm$^{-2}$s$^{-1}$. 
%This allows measuring muon momentum with ... momentum resolution.
Spacial and momentum information about muons in the MS is used to match them to tracks in the vertex spectrometer with high efficiency and minimal false-positive rates.
By matching tracks, in coordinate and momentum space, it is possible to accurately measure the muon kinematics, reaching a resolution of less than $10$\,MeV at the $\omega$-meson mass and $\sim 30$\,MeV at the J/$\psi$. Moreover, the high granularity of the vertex detectors allows for accurate measurements of hadronic decays of open charm and strange hadrons.
An additional 180\,cm thick graphite absorber follows the MS. A pair of pad stations, similar in technology to the chambers of the MS, can be used as a trigger and/or as a muon identifier in the experiment. 

\begin{figure}[h]
    \centering
      \includegraphics[width=.43\textwidth]{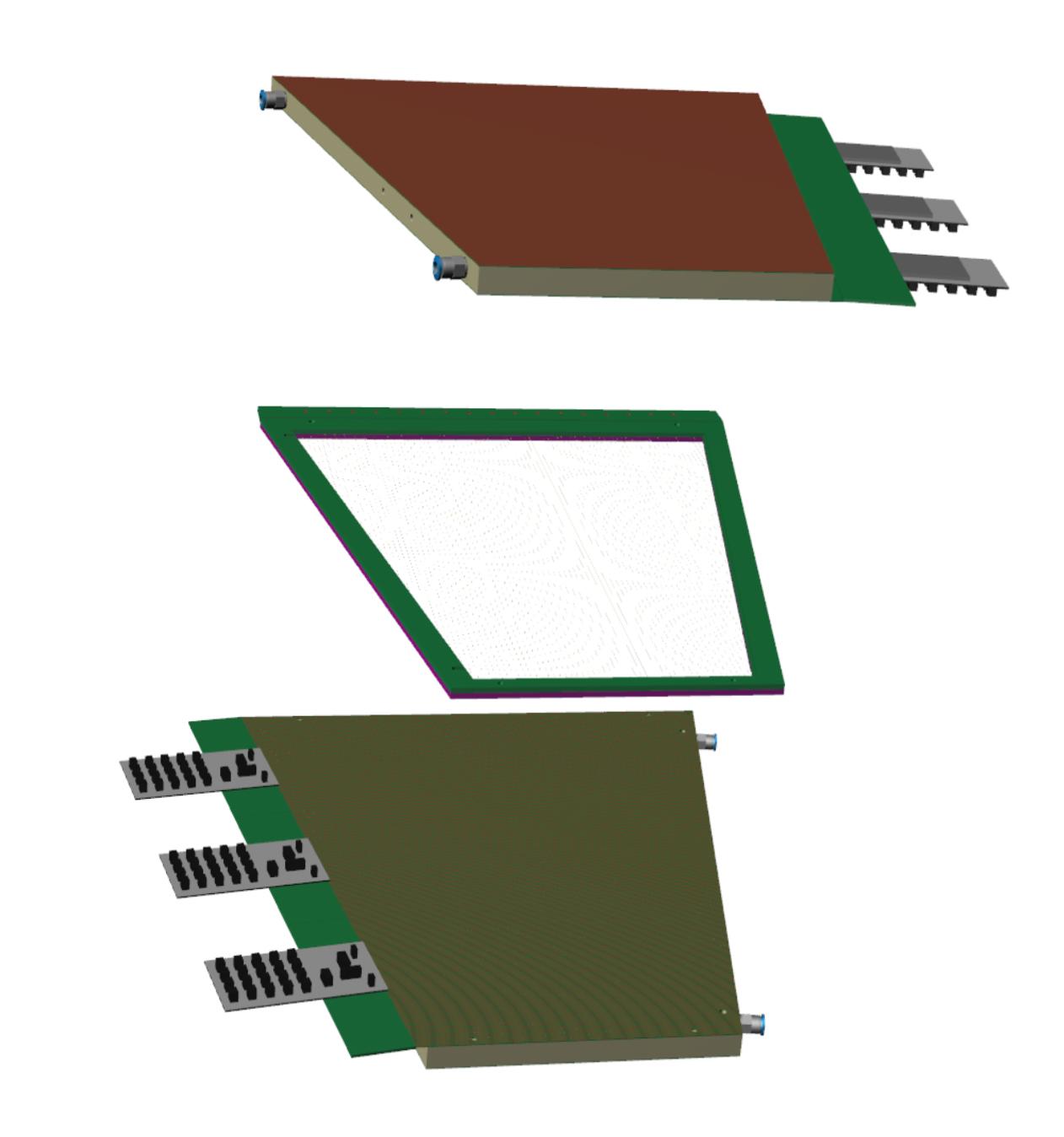}
      \includegraphics[width=.43\textwidth]{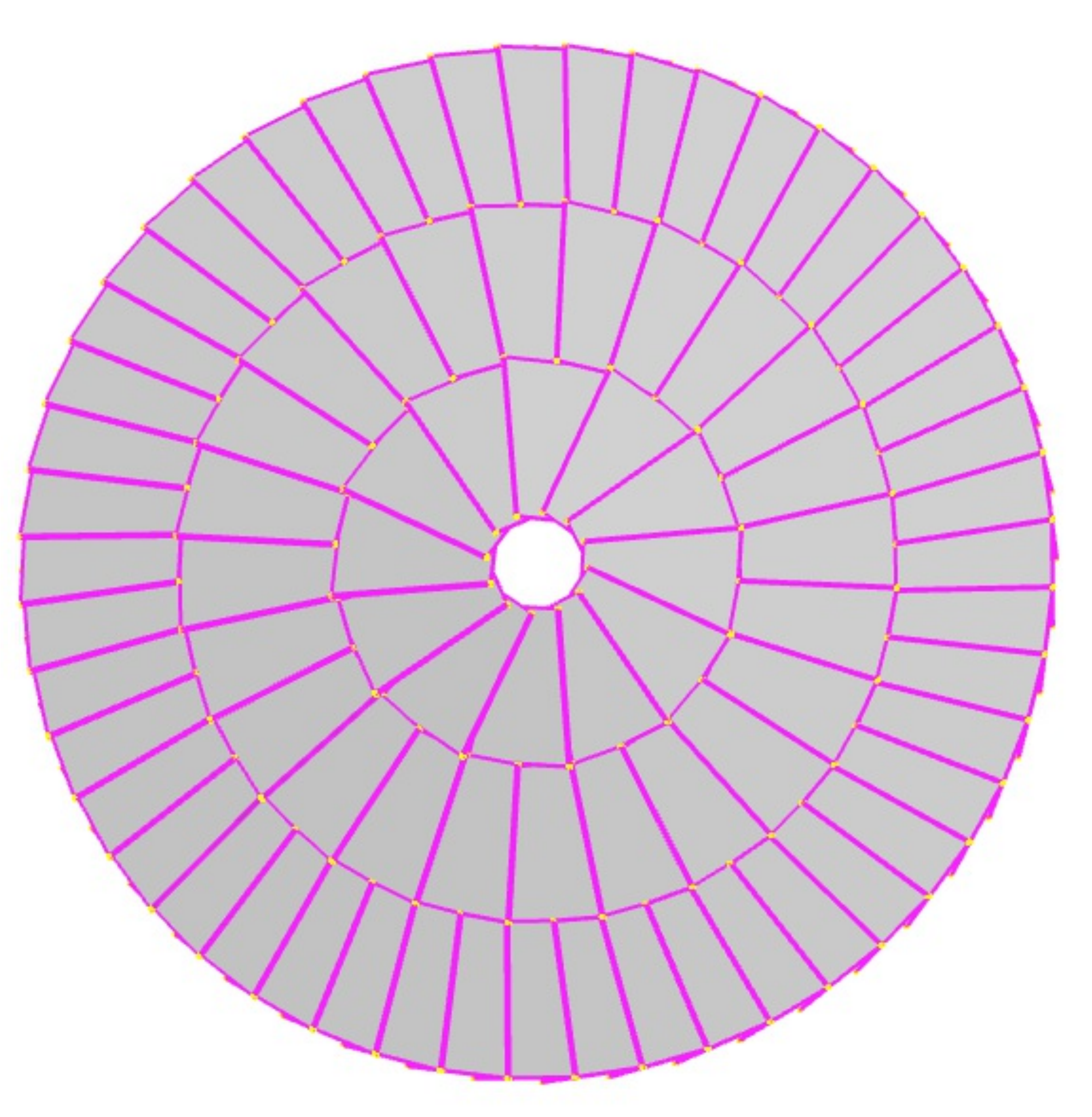}
    \caption{The exploded view of the MWPC detector prototype and the CAD rendering of the largest muon tracking stations (right). }
    \label{fig:wheels}
\end{figure}

An R\&D program is in progress for defining the detector aspects. For the MAPS, the studies are advancing in the frame of a collaboration with ALICE, intending to produce, via a stitching technique, sufficient enough surface of detectors with a low-material budget.
For the MWPC, the first prototype has been built, and it is being tested at the detector construction facility at the Weizmann Institute of Science and is planned to be exposed to an SPS test beam. The tracking stations are designed to be built by replicating the developed prototype to obtain the desired geometry. The two upstream muon stations are composed of 12 trapezoid detector modules each, while those downstream of the toroidal magnet and the cylindrical graphite absorber have 36 and 84 modules, respectively. Figure~\ref{fig:wheels} shows the design of the MWPC module (left) and the foreseen arrangement of the modules for the largest stations (right).
For the toroidal magnet, a scaled (1:5) working prototype has been built and tested to assess the feasibility of producing a device with the needed geometry and to provide the necessary information to design the full-scale device.  

%Measurements with a Pb beam need to be complemented by corresponding data-taking periods with a proton beam incident on various nuclear targets, collecting an equivalent luminosity and providing reference data for the correct quantitative interpretation of nuclear collision results.

%The primary function of the muon spectrometer is to identify muons among other particles, measure their momentum, and match them to the vertex telescope. This is done by reconstructing charged particle tracks after the absorber and measuring their curvature in the magnetic field of the toroidal magnet.
% The muon tracker provides the track's momentum by measuring the difference in the polar angle between the first and the second pairs of layers. 

%\section{Summary}

The NA60+ project, thanks to its rich and specific physics program, will play a unique role by providing for the first time accurate experimental data on dileptons, open charm, and charmonia within the range of SPS energies.
The running plan of the experiment for the first 5-6 years includes monthly periods with a Pb beam, with different incident energy for each year, complemented by a few weeks of proton beam running for calibration purposes and physics studies with $p$-A collisions. This will ensure a fine enough energy scan for the 
characterization of the QGP at varying $\mu_{\rm B}$ and the search of signals related to the first-order phase transition.
The proposed new experiment is based on state-of-the-art technologies that would allow the collection of about 20 times larger statistics at each energy, with respect to the former NA60 experiment which only operated at top SPS energy. 
Despite the challenging conditions, the experimental layout is well adapted to the physics tasks of the NA60+ experiment. 

%\acknowledgments

\end{document}